\documentclass{llncs}%
\usepackage{amsmath}
\usepackage{amssymb}
\usepackage{amsfonts}
\usepackage{makeidx}
\usepackage{url}%
\usepackage{graphicx}
\providecommand{\U}[1]{\protect\rule{.1in}{.1in}}
\input{Qcircuit}
\def\rajaack{\thanks{R. Nagarajan is supported by EPSRC grant GR/S34090
and the EU Sixth Framework Programme (Project SecoQC: \textit
{Development of a Global Network for Secure Communication based on Quantum Cryptography}).}}
\def\nickack{\thanks{N. Papanikolaou is supported by a Postgraduate Fellowship
by the Department of Computer Science, University of Warwick.}}
\def\lessfltspace{-6mm}
\def\lessmathspace{-7.5mm}
\makeatletter
\renewcommand{\paragraph}{\@startsection{paragraph}{4}{0em}{0.4\baselineskip}{-\fontdimen2\font plus -\fontdimen3\font minus -\fontdimen4\font}{\normalfont\normalsize\itshape}}
\makeatother
\begin{document}

\frontmatter
\title{Probabilistic Model--Checking\\of Quantum Protocols }%

%TCIMACRO{\TeXButton{\mainmatter}{\mainmatter}}%
%BeginExpansion
\mainmatter
%EndExpansion%
%TCIMACRO{\TeXButton{Authors}{\def\aj{a{\kern-0.2pt}j} \author{Simon Gay\inst
%{1} \and R\aj agopal Nagar\aj an\inst{2}$^{,}$\rajaack\and
%Nikolaos Papanikolaou\inst{2}$^{,}$\nickack}}}%
%BeginExpansion
\def\aj{a{\kern-0.2pt}j} \author{Simon Gay\inst{1} \and R\aj agopal Nagar\aj
an\inst{2}$^{,}$\rajaack\and Nikolaos Papanikolaou\inst{2}$^{,}$\nickack}%
%EndExpansion
%

%TCIMACRO{\TeXButton{Author addresses}{\institute{
%Department of Computing Science,
%University of Glasgow\\
%\email{simon@dcs.gla.ac.uk}
%\and Department of Computer Science, University of Warwick\\
%\email{\{biju,nikos\}@dcs.warwick.ac.uk}
%}}}%
%BeginExpansion
\institute{
Department of Computing Science,
University of Glasgow\\
\email{simon@dcs.gla.ac.uk}
\and Department of Computer Science, University of Warwick\\
\email{\{biju,nikos\}@dcs.warwick.ac.uk}
}%
%EndExpansion%
%TCIMACRO{\TeXButton{Make title}{\maketitle}}%
%BeginExpansion
\maketitle
%EndExpansion
%

%TCIMACRO{\TeXButton{Begin abstract}{\begin{abstract}}}%
%BeginExpansion
\begin{abstract}%
%EndExpansion
We establish fundamental and general techniques for formal verification of
quantum protocols. Quantum protocols are novel communication schemes involving
the use of quantum-mechanical phenomena for representation, storage and
transmission of data. As opposed to quantum computers, quantum communication
systems can and have been implemented using present-day technology; therefore,
the ability to model and analyse such systems rigorously is of primary importance.

While current analyses of quantum protocols use a traditional mathematical
approach and require considerable understanding of the underlying physics, we
argue that automated verification techniques provide an elegant alternative.
We demonstrate these techniques through the use of \textsc{prism}, a
probabilistic model-checking tool. Our approach is conceptually simpler than
existing proofs, and allows us to disambiguate protocol definitions and assess
their properties. It also facilitates detailed analyses of actual implemented
systems. We illustrate our techniques by modelling a selection of quantum
protocols (namely superdense coding, quantum teleportation, and quantum error
correction) and verifying their basic correctness properties. Our results
provide a foundation for further work on modelling and analysing larger
systems such as those used for quantum cryptography, in which basic protocols
are used as components.%

%TCIMACRO{\TeXButton{End abstract}{\end{abstract}}}%
%BeginExpansion
\end{abstract}%
%EndExpansion

\section{Introduction}

In the 1980s Richard Feynman, David Deutsch, Paul Benioff, and other
scientists realized that quantum--mechanical phenomena can be exploited
directly for the manipulation, storage and transmission of information. The
discovery of quantum algorithms for prime factorization \cite{Shor94} and
unstructured search \cite{Grover-96}, which outperformed the best classical
algorithms for these tasks significantly, opened up new vistas for computer
science and gave an initial thrust to the emerging field of quantum
computation. To implement a quantum algorithm, however, a large scale quantum
computer is necessary and such a device has yet to be built. Research in
\emph{quantum information,} on the other hand, has shown that quantum effects
can be harnessed to provide efficient and highly secure communication
channels, which can be built using current technology. Entangled quantum
states, superpositions and quantum measurement are among the characteristics
of the subatomic world which nature puts at our disposal; these and related
phenomena enable the development of novel techniques for computation and
communication with no rival in classical computing and communication theory.

The focus in this paper is on communication protocols involving the use of
such phenomena. Quantum protocols have particularly important applications in
cryptography. Several quantum protocols have been proposed for cryptographic
tasks such as oblivious transfer, bit commitment and key distribution
\cite{gruska,nielsenchuang}. The BB84 protocol for quantum key distribution
\cite{bb84-orig,nikos-acmCR}, which allows two users to establish a common
secret key using a single quantum channel, has been shown to be
unconditionally secure against all attacks \cite{Mayers01}. This degree of
security has never been guaranteed by any classical cryptographic protocol,
and the discovery has incited widespread interest in the properties of quantum
protocols. Furthermore, practical quantum cryptographic devices are
commercially available (e.g. from the companies Id\ Quantique and MagiQ).

The quantum teleportation protocol \cite{Bennett93} and the superdense coding
protocol \cite{Bennett92} are both interesting and important examples of
non--cryptographic quantum communication. Also, any attempt to implement a
realistic quantum communication system, or indeed a quantum computer, must
account for noise and quantum decoherence; these phenomena may be tackled
through the use of quantum error correction protocols \cite{steane-err-corr}.
We will study these protocols in detail.

We argue that detailed, automated analyses of protocols such as these
facilitate our understanding of complex quantum behaviour and enable us to
construct valuable proofs of correctness. Such analyses are especially
important to manufacturers of commercial devices based on such protocols; the
actual security of commercial quantum cryptographic systems, for example, is
worth an in--depth investigation. Communication protocols have always been
under scrutiny by computer scientists, who have developed numerous techniques
for analysing and testing them, including process algebras, formal
specification languages and automated verification tools. Automated
verification techniques, such as \emph{model-checking} and \emph{theorem
proving,} are frequently targeted at protocols and have been used to detect
faults and subtle bugs. For instance, the \textsc{fdr} model-checker allowed
Gavin Lowe to uncover a flaw in the Needham--Schroeder security protocol
\cite{pink-protocols}.

Our approach is distinguished by the ability to incorporate system parameters
in models and to vary them during verification. Also, while manual proofs of
correctness have to be rewritten to account for variations in a protocol, the
models we use may be adapted easily to different scenarios. Although current
model-checkers were developed primarily for the analysis of
classical\ systems, we have found ways of using them to model quantum
behaviour. To account for the probabilism inherent in quantum systems, we have
chosen to use a \emph{probabilistic} model--checker, in particular, the
\textsc{prism} tool developed at the University of Birmingham
\cite{prism-manual}.

\ The structure of this paper is as follows. Section 2 provides necessary
background on quantum theory and the \textsc{prism} model checker. In Section
3, we discuss the issues associated with modelling quantum protocols, and
explain how they may be resolved through explicit state space generation. This
leads to a presentation of specific quantum protocols, namely superdense
coding, quantum teleportation, and the quantum bit-flip code. We explain how
the techniques of Section 3 have enabled us to analyse and validate these
protocols with \textsc{prism}. Section 5 sets the stage for future work.
Finally, we conclude with the conviction that our techniques may be applied to
more substantial systems, containing both classical and quantum components.%

%TCIMACRO{\TeXButton{Previous and Related Work}{\paragraph
%{Previous and Related Work.}} }%
%BeginExpansion
\paragraph{Previous and Related Work.}
%EndExpansion
Formal verification of quantum protocols was advocated by Nagarajan and Gay in
\cite{rajaverif}, where they use the process calculus \textsc{ccs} to formally
specify BB84. Papanikolaou's M.Sc. thesis \cite{nikosthesis} details a
preliminary analysis of the BB84 quantum key distribution protocol using the
\textsc{prism} and the \textsc{spin} model checkers.%

%TCIMACRO{\TeXButton{Acknowledgement}{\paragraph{Acknowledgement.}}}%
%BeginExpansion
\paragraph{Acknowledgement.}%
%EndExpansion
The authors would like to thank Marta Kwiatkowska's group at the University of
Birmingham, especially David Parker and Gethin Norman, for assistance with the
\textsc{prism} tool.

\section{Preliminaries}

\subsection{Basic Concepts of Quantum Computation}

We briefly introduce those aspects of quantum theory relevant to quantum
protocols. More detailed presentations can be found in \cite{nielsenchuang}
and \cite{rieffel}.

A \emph{quantum bit} or \emph{qubit} is a physical system which has two basis
states, conventionally written $|0\rangle$ and $|1\rangle$, corresponding to
one-bit classical values. These could be, for example, spin states of a
particle or polarization states of a photon, but we do not consider physical
details. According to quantum theory, a general state of a quantum system is a
\emph{superposition} or linear combination of basis states. A qubit has state
$\alpha|0\rangle+\beta|1\rangle$, where $\alpha$ and $\beta$ are complex
numbers such that $|{\alpha}|^{2}+|{\beta}|^{2}=1$; states which differ only
by a (complex) scalar factor with modulus $1$ are indistinguishable. States
can be represented by column vectors: $%
%TCIMACRO{\QATOPD{[}{]}{\alpha}{\beta}}%
%BeginExpansion
\genfrac{[}{]}{0pt}{}{\alpha}{\beta}%
%EndExpansion
=\alpha|0\rangle+\beta|1\rangle.$ Formally, a quantum state is a unit vector
in a Hilbert space, i.e.\ a complex vector space equipped with an inner
product satisfying certain axioms. In this paper we restrict attention to
collections of qubits.

The basis $\{|0\rangle,|1\rangle\}$ is known as the \emph{standard} basis.
Other bases are sometimes of interest, especially the \emph{diagonal} (or
\emph{dual}, or \emph{Hadamard}) basis consisting of the vectors
\[
|+\rangle=\frac{1}{\sqrt{2}}(|0\rangle+|1\rangle)\text{ \ \ and \ }%
|-\rangle=\frac{1}{\sqrt{2}}(|0\rangle-|1\rangle)
\]

Evolution of a closed quantum system can be described by a \emph{unitary
transformation}. If the state of a qubit is represented by a column vector
then a unitary transformation $U$ can be represented by a complex-valued
matrix $(u_{ij})$ such that $U^{-1}=U^{\ast}$, where $U^{\ast}$ is the
conjugate-transpose of $U$ (i.e.\ element $ij$ of $U^{\ast}$ is $\bar{u}_{ji}%
$). $U$ acts by matrix multiplication:
\[%
\begin{bmatrix}
{\alpha^{\prime}}\\
{\beta^{\prime}}%
\end{bmatrix}
=%
\begin{bmatrix}
{u_{00}} & {u_{01}}\\
{u_{10}} & {u_{11}}%
\end{bmatrix}%
\begin{bmatrix}
{\alpha}\\
{\beta}%
\end{bmatrix}
\]
A unitary transformation can also be defined by its effect on basis states,
which is extended linearly to the whole space. For example, the
\emph{Hadamard} operator is defined by
\[%
\begin{array}
[c]{lclclcl}%
|0\rangle & \mapsto & |+\rangle=\frac{1}{\sqrt{2}}|0\rangle+\frac{1}{\sqrt{2}%
}|1\rangle & \hspace{10mm} & |1\rangle & \mapsto & |-\rangle=\frac{1}{\sqrt
{2}}|0\rangle-\frac{1}{\sqrt{2}}|1\rangle
\end{array}
\]
which corresponds to the matrix $\mathsf{H}=\frac{1}{\sqrt{2}}%
\begin{bmatrix}
{1} & {1}\\
{1} & {-1}%
\end{bmatrix}
$. The \emph{Pauli} operators, denoted by $\sigma_{0},\sigma_{1},\sigma
_{2},\sigma_{3}$, are defined by
%\[
%\begin{array}{@{\extracolsep{2mm}}cccc}
%\text{$I$ or $\sigma_0$} & \text{$\sigma_x$ or $\sigma_1$} & \text{$\sigma_y$ or $\sigma_2$} & \text{$\sigma_z$ or $\sigma_3$} \\ \\
%\matr{1}{0}{0}{1} & \matr{0}{1}{1}{0} &
%\matr{0}{-i}{i}{0} & \matr{1}{0}{0}{-1}
%\end{array}
%\]%
\[%
\begin{array}
[c]{c@{\extracolsep{5mm}}ccc}%
\sigma_{0}=%
\begin{bmatrix}
{1} & {0}\\
{0} & {1}%
\end{bmatrix}
& \sigma_{1}=%
\begin{bmatrix}
{0} & {1}\\
{1} & {0}%
\end{bmatrix}
& \sigma_{2}=%
\begin{bmatrix}
{0} & {-i}\\
{i} & {0}%
\end{bmatrix}
& \sigma_{3}=%
\begin{bmatrix}
{1} & {0}\\
{0} & {-1}%
\end{bmatrix}
\end{array}
\]

Measurement plays a key role in quantum physics. If a qubit is in state
$\alpha|0\rangle+\beta|1\rangle$ then measuring its value gives the result $0$
with probability $|{\alpha}|^{2}$ (leaving it in state $|0\rangle$) and the
result $1$ with probability $|{\beta}|^{2}$ (leaving it in state $|1\rangle
$).
%Protocols sometimes specify measurement with respect to a
%different basis, such as the diagonal basis; this can be expressed as
%a unitary change of basis followed by a measurement with respect to
%the standard basis.
For example, if a qubit is in state $|+\rangle$ then a measurement (with
respect to the standard basis) gives result $0$ (and state $|0\rangle$) with
probability $\frac{1}{2}$, and result $1$ (and state $|1\rangle$) with
probability $\frac{1}{2}$. If a qubit is in state $|0\rangle$ then a
measurement gives result $0$ (and state $|0\rangle$) with probability $1$.
%result\footnote{Strictly speaking, the outcome of the measurement is
%just the final state; the specific association of numerical results
%with final states is a matter of convention.} $0$ (and state
%$\ket{+}$) with probability $\frac{1}{2}$, and result $1$ (and state
%$\ket{-})$) with probability $\frac{1}{2}$, because of the
%representation of $\ket{0}$ in the diagonal basis noted above. If a
%classical bit is represented by a qubit using either the standard or
%diagonal basis, then a measurement with respect to the correct basis
%results in the original bit, but a measurement with respect to the
%other basis results in $0$ or $1$ with equal probability. This
%behaviour is used by the quantum bit-commitment protocol which we
%discuss in Section~\ref{sec-bitcommitment}.

To go beyond single-qubit systems, we consider tensor products of spaces (in
contrast to the cartesian products used in classical systems). If spaces $U$
and $V$ have bases $\{u_{i}\}$ and $\{v_{j}\}$ then $U\otimes V$ has basis
$\{u_{i}\otimes v_{j}\}$. In particular, a system consisting of $n$ qubits has
a $2^{n}$-dimensional space whose standard basis is $|00\ldots0\rangle
\ldots|11\ldots1\rangle$. We can now consider measurements of single qubits or
collective measurements of multiple qubits. For example, a $2$-qubit system
has basis $|00\rangle,|01\rangle,|10\rangle,|11\rangle$ and a general state is
$\alpha|00\rangle+\beta|01\rangle+\gamma|10\rangle+\delta|11\rangle$ with
$|{\alpha}|^{2}+|{\beta}|^{2}+|{\gamma}|^{2}+|{\delta}|^{2}=1$. Measuring the
first qubit gives result $0$ with probability $|{\alpha}|^{2}+|{\beta}|^{2}$
(leaving the system in state $\frac{1}{\sqrt{|{\alpha}|^{2}+|{\beta}|^{2}}%
}(\alpha|00\rangle+\beta|01\rangle)$) and result $1$ with probability
$|{\gamma}|^{2}+|{\delta}|^{2}$ (leaving the system in state $\frac{1}%
{\sqrt{|{\gamma}|^{2}+|{\delta}|^{2}}}(\gamma|10\rangle+\delta|11\rangle)$);
in each case we renormalize the state by multiplying by a suitable scalar
factor. Measuring both qubits simultaneously gives result $0$ with probability
$|{\alpha}|^{2}$ (leaving the system in state $|00\rangle$), result $1$ with
probability $|{\beta}|^{2}$ (leaving the system in state $|01\rangle$) and so
on; the association of basis states $|00\rangle,|01\rangle,|10\rangle
,|11\rangle$ with results $0,1,2,3$ is just a conventional choice. The power
of quantum computing, in an algorithmic sense, results from calculating with
superpositions of states; all of the states in the superposition are
transformed simultaneously (\emph{quantum parallelism}) and the effect
increases exponentially with the dimension of the state space. The challenge
in quantum algorithm design is to make measurements which enable this
parallelism to be exploited; in general this is very difficult.

The \emph{controlled not} ($\mathsf{CNot}$) operator on pairs of qubits
performs the mapping $|00\rangle\mapsto|00\rangle$, $|01\rangle\mapsto
|01\rangle$, $|10\rangle\mapsto|11\rangle$, $|11\rangle\mapsto|10\rangle$,
which can be understood as inverting the second qubit (the \emph{target}) if
and only if the first qubit (the \emph{control}) is set. The action on general
states is obtained by linearity.

Systems of two or more qubits may be in \emph{entangled} states, meaning that
the states of the qubits are correlated. For example, consider a measurement
of the first qubit of the state $\frac{1}{\sqrt{2}}(|00\rangle+|11\rangle)$.
The result is $0$ (and the resulting state is $|00\rangle$) with probability
$\frac{1}{2}$, or $1$ (and the resulting state is $|11\rangle$) with
probability $\frac{1}{2}$. In either case, a subsequent measurement of the
second qubit gives a definite, non--probabilistic result which is identical to
the result of the first measurement. This is true even if the entangled qubits
are physically separated. Entanglement illustrates the key difference between
the use of the tensor product (in quantum systems) and the cartesian product
(in classical systems): an entangled state of two qubits is one which cannot
be expressed as a tensor product of single-qubit states. The Hadamard and
$\mathsf{CNot}$ operators can be combined to create entangled states:
$\mathsf{CNot}((\mathsf{H}\otimes I)|00\rangle)=\frac{1}{\sqrt{2}}%
(|00\rangle+|11\rangle)$.

\subsection{Probabilistic Model--Checking\label{sec:probmc}}

\textsc{prism} is an acronym for \emph{probabilistic symbolic model checker,}
and is designed for modelling and validating systems which exhibit
probabilistic behaviour. Whereas a logical model--checker, such as
\textsc{spin} \cite{spin-newbook}, only states whether a system model $\sigma$
satisfies a temporal formula $\Phi$, a tool such as \textsc{prism} computes
the probability with which such a formula is satisfied, i.e. the value of
$\mathrm{P}_{\sigma,\Phi}=\Pr\{\sigma\models\Phi\}$ \noindent for given
$\sigma$ and $\Phi$. The models catered for by \textsc{prism} may incorporate
specific probabilities for various behaviors and so may the formulas used for
verification. Probabilistic models and \textsc{prism}--like tools find
applications in numerous areas of computer science where random behaviour is
involved. Oft--cited applications are randomized algorithms, real--time
systems and Monte Carlo simulation. The application of probabilistic
model--checking to quantum systems is entirely appropriate, since quantum
phenomena are inherently described by random processes; to reason about such
phenomena one must account for this.

\textsc{prism} uses a built--in specification language based on Alur and
Henzinger's \textsc{reactive modules} formalism (see
\cite{Kwiatkowska04,prism-manual} for details). Using this language the user
can describe probabilistic behaviour. Internally, a \textsc{prism} model is
represented by a \emph{probabilistic transition system.} In such a system,
each step in a computation is represented by a \emph{move,} or
\emph{transition,} from a particular state $s$ to a distribution $\pi$ of
successor states. For technical details, refer to \cite{prism-manual}.

The probabilistic temporal logic \textsc{pctl} \cite{CiesinskiPCTL}\ is used
as the principal means for defining properties of systems modelled in
\textsc{prism}. It suffices for our purposes to remind the reader of the
meaning of the operator $\mathcal{U}$, known as \textquotedblleft unbounded
until\textquotedblright. The formula $\Phi_{1}\,\mathcal{U}\,\Phi_{2}$
expresses the fact that $\Phi_{1}$ holds continuously from the current state
onward, \emph{until eventually} $\Phi_{2}$ becomes $\mathbf{true}$. The
\textsc{prism} property $\mathrm{P}\geqslant1[\Phi_{1}\,\mathcal{U}\,\Phi
_{2}]$ states that the formula $\Phi_{1}\,\mathcal{U}\,\Phi_{2}$ is true with
certainty, i.e. with a probability of unity; we use \textsc{prism} to check
whether such a property holds in a given model.

\section{Fundamental Techniques}

In order to use a classical probabilistic model--checker to verify quantum
protocols, we need to model the quantum states that arise in a given protocol,
and the effect of specific quantum operations on these states. \textsc{prism}
itself only allows positive integer and boolean variables to be used in
models.\ So how can we model the states of quantum systems, and the quantum
operations arising in protocols, using only classical data types and arithmetic?

Single qubits can be in a superposition of two states, while classical
variables can only take on a single value in any given state. The coefficients
of these states can be any two complex numbers whose moduli squared sum to
unity, and there is an uncountable infinity of these; of course,
\textsc{prism} can only work with a finite state space. Furthermore, quantum
systems consisting of many qubits can be in entangled states, which, unlike
classical systems, cannot be decomposed into products of individual states.
What is needed, therefore, is a means of representing quantum states fully and
consistently, in a form that \textsc{prism} can handle.

Of all the possible quantum states of an $n$--qubit system, we identify the
finite set of states which arise by applying the operations $\mathsf{CNot}$,
Hadamard ($H$), and $\sigma_{0},\sigma_{1},\sigma_{2},\sigma_{3}$ to input
states. We confine our analyses to protocols that involve \emph{only} this
restricted set of operations. At present, determining which states belong to
this set is done manually, but we are considering ways of automating this.

Consider a very simple system: a single qubit, being acted upon through the
Hadamard gate and through measurement in the standard basis. For our purposes,
the state of the qubit may be $\left\vert 0\right\rangle ,\left\vert
1\right\rangle ,$ or an equal superposition of the two. In fact, these states
are sufficient to model the BB84 protocol for quantum key distribution
\cite{bb84-orig}. The quantum states which we need to represent in order to
model this simple system are thus:
\[
\left\vert 0\right\rangle ,\left\vert 1\right\rangle ,\frac{1}{\sqrt{2}%
}\left(  \left\vert 0\right\rangle +\left\vert 1\right\rangle \right)  ,\text{
and }\frac{1}{\sqrt{2}}\left(  \left\vert 0\right\rangle -\left\vert
1\right\rangle \right)
\]

\noindent To model this small, finite set of quantum states, which is closed
under the operation of the Hadamard gate and the Pauli operators, we represent
each state by assigning a unique integer from 0 to 3 to it, and we use
straightforward transitions from one integer value to another to model the
action of the gate. The actual \textsc{prism} model for this, as well as the
possible results of measurement, is listed in the appendix.

A protocol such as superdense coding, which we will discuss in Section
\ref{sec:densecod}, can be expressed as a step-by-step interaction with a
\emph{two--qubit system. }In order to model the states of 2-- and 3--qubit
systems, the quantum operators and the measurements which arise in this and
related protocols such as teleportation, we have developed a code generation
tool called \textsc{prismgen}. This tool generates a \textsc{prism} code
fragment, or \emph{module,} in which each quantum state is represented by a
unique positive integer. Every quantum operator used in a particular protocol
is coded as a set of deterministic transitions from one quantum state to
another. \textsc{prismgen} calculates these transitions by multiplying the
unitary matrix, which corresponds to a particular operator, with each quantum
state vector of interest. A measurement is modelled by a set of probabilistic
transitions, leading to the various possible outcomes with equal probability.
For simplicity, we have only considered states whose measurement outcomes are
all equiprobable, although \textsc{prism} does allow us to model the more
general case.

From the overall state space for a two--qubit system, a certain subset is
closed under the $\mathsf{CNot}$, Hadamard and Pauli operations. This subset
consists of the following states, which are 24 in total:

\begin{itemize}
\item 4 states corresponding to the four basis vectors, i.e. $\left\vert
00\right\rangle ,\left\vert 01\right\rangle ,\left\vert 10\right\rangle
,\left\vert 11\right\rangle .$

\item 12 states which are sums of two basis vectors, i.e. of the form
$\frac{1}{\sqrt{2}}\left\vert xy\right\rangle \pm\frac{1}{\sqrt{2}}\left\vert
x^{\prime}y^{\prime}\right\rangle $ with $x\neq x^{\prime},y\neq y^{\prime}$
and $x,x^{\prime},y,y^{\prime}\in\{0,1\}.$

\item 8 states which are sums of all four basis vectors.
\end{itemize}%

%TCIMACRO{\TeXButton{Proposition.}{\paragraph{Proposition.}} }%
%BeginExpansion
\paragraph{Proposition.}
%EndExpansion
The above set of 24 states is closed under the $\mathsf{CNot}$, Hadamard and
Pauli operations.

\begin{proof}
{\small These states can be expressed in the following way.}
\end{proof}

\begin{enumerate}
\item {\small The single basis vectors: }$|00\rangle${\small , }$|01\rangle
${\small , }$|10\rangle${\small , }$|11\rangle$

\item {\small The states containing two basis vectors can be separated into
three subclasses:}

\begin{enumerate}
\item $\frac{1}{\sqrt{2}}\left(  |0\rangle\pm|1\rangle\right)  \otimes
|0\rangle$, $\ \frac{1}{\sqrt{2}}\left(  |0\rangle\pm|1\rangle\right)
\otimes|1\rangle$

\item $|0\rangle\otimes\frac{1}{\sqrt{2}}\left(  |0\rangle\pm|1\rangle\right)
$, $|1\rangle\otimes\frac{1}{\sqrt{2}}\left(  |0\rangle\pm|1\rangle\right)  $

\item $\frac{1}{\sqrt{2}}(|00\rangle\pm|11\rangle)$, $\frac{1}{\sqrt{2}%
}(|01\rangle\pm|10\rangle)$
\end{enumerate}

\item {\small The states containing four basis vectors can be expressed in any
of the forms:}

\begin{enumerate}
\item $\frac{1}{\sqrt{2}}\left(  |0\rangle\pm|1\rangle\right)  \otimes
|0\rangle\,\pm\,\frac{1}{\sqrt{2}}\left(  |0\rangle\pm|1\rangle\right)
\otimes|1\rangle$

\item $|0\rangle\otimes\frac{1}{\sqrt{2}}\left(  |0\rangle\pm|1\rangle\right)
\,\pm\,|1\rangle\otimes\frac{1}{\sqrt{2}}\left(  |0\rangle\pm|1\rangle\right)
$

\item $\frac{1}{2}(|00\rangle\,\pm\,|01\rangle\,\pm\,|10\rangle\,\pm
\,|11\rangle)$
\end{enumerate}
\end{enumerate}

{\small It is obvious that each set (1.)--(3.) individually is closed under
each }$\sigma_{i}${\small \ (applied to either qubit) and }$\mathsf{CNot}$
%{\small \ (with either qubit as control) because
these operations are permutations of the basis vectors. Each set has an
evident symmetry among the basis vectors (taking (3.a) and expanding (2.) into
an explicit list of states). Applying $H${\small \ to the first qubit gives a
bijection between (1.) and (2.a), between (2.b) and (3.a), and between (2.c)
and (3.b). Applying }$H${\small \ to the second qubit is similar. \hfill
}$\square$

~

Our \textsc{prismgen} tool enumerates these states and calculates the
transitions corresponding to the various operations. The resulting
\textsc{prism} module can be included as part of any model which involves
measurements and the application of these operations to a system of two qubits.

The situation with a system of three qubits is similar. We have developed a
3--qubit version of \textsc{prismgen,} which gives us the ability to model
protocols such as those for quantum teleportation and quantum error correction.

\section{Illustrative Examples\label{sec:examples}}

We have been able to model a certain number of quantum protocols using the
aforementioned techniques. These include: (1) superdense coding, which is a
procedure for encoding pairs of classical bits into single qubits; (2) quantum
teleportation, which allows the transmission of a quantum state without the
use of an intervening quantum channel; and (3) quantum error correction,
namely the qubit flip code, which corrects a single bit flip error during
transmission of quantum bits. The source files for the models in this section
will be made available online at
%TCIMACRO{\TeXButton{Nikos' URL}{\url{http://go.warwick.ac.uk/nikos/research/}%
%}}%
%BeginExpansion
\url{http://go.warwick.ac.uk/nikos/research/}%
%EndExpansion
.

\subsection{Superdense Coding\label{sec:densecod}}

The simplest quantum protocol which we will use to illustrate our techniques
is the superdense coding scheme \cite{Bennett92}. This scheme makes it
possible to encode a pair of classical bits on a single qubit. With superdense
coding, a quantum channel with a capacity of a single qubit is all that is
necessary to transmit twice as many bits as a serial classical channel.
Superdense coding is essentially a computation on a two--qubit system;
therefore, the \textsc{prism} model of this protocol uses the 2--qubit version
of \textsc{prismgen}. We begin with a description of the protocol, and proceed
to show how it is modelled and verified with \textsc{prism}.

\begin{figure}[ht]
\centerline{\Qcircuit@C=1em @R=0.8em {
& \\
& \\
\lstick{i=2x+y}   & \cw& \cw& \cw& \cw&  \control\cw\cwx[1]            \\
\lstick{P_1:  \ket{0}} & \qw& \gate{H}  & \ctrl{2}  &\qw& \gate{\sigma_i}
& \qw& \ctrl{2} & \gate{H} & \qw\qwx[1]  \\
&         &                 &               &         &                           &        &              &               &                      & \qw
&\meter&\cw&\rstick{x'}\cw\\
\lstick{P_2:  \ket{0}} & \qw& \qw& \targ&\qw&\qw& \qw& \targ& \qw& \qw
& \qw&\meter&\cw&\rstick{y'}\cw& \\
}}
\caption{Quantum circuit diagram for the superdense coding protocol.}
\label{fig:dc-circ}
\end{figure}
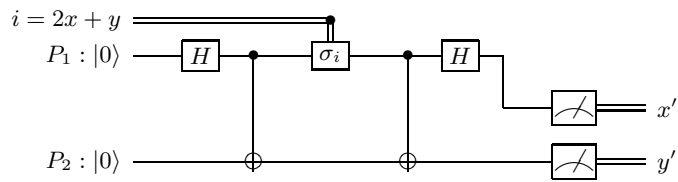

The setting for superdense coding involves two parties, conventionally named
Alice and Bob, who are linked by a quantum channel and share a pair of
entangled qubits. The objective is for Alice to communicate the binary number
$xy$ --- henceforth termed the \emph{message} and denoted by $(x,y)$, with
$x,y\in\{0,1\}$ --- by transmitting a single qubit to Bob. The superdense
protocol takes advantage of the correlations between qubits $P_{1} $ and
$P_{2}$, which are in an entangled quantum state. Alice essentially influences
this state in such a way that Bob's measurement outcome matches the message of
her choice. The quantum circuit diagram for the superdense coding procedure is
shown in Fig. \ref{fig:dc-circ}.%

%TCIMACRO{\TeXButton{FIGURE: Superdense coding FULL circuit}{\begin{figure}[ht]
%\centerline{\Qcircuit@C=1em @R=0.8em {
%& \\
%& \\
%\lstick{i=2x+y}   & \cw& \cw& \cw& \cw&  \control\cw\cwx[1]            \\
%\lstick{P_1:  \ket{0}} & \qw& \gate{H}  & \ctrl{2}  &\qw& \gate{\sigma_i}
%& \qw& \ctrl{2} & \gate{H} & \qw\qwx[1]  \\
%&         &                 &               &         &                           &        &              &               &                      & \qw
%&\meter&\cw&\rstick{x'}\cw\\
%\lstick{P_2:  \ket{0}} & \qw& \qw& \targ&\qw&\qw& \qw& \targ& \qw& \qw
%& \qw&\meter&\cw&\rstick{y'}\cw& \\
%}}
%\caption{Quantum circuit diagram for the superdense coding protocol.}
%\label{fig:dc-circ}
%\end{figure}}}%
%BeginExpansion
%EndExpansion

Quantum circuit diagrams are a convenient means for expressing computations on
qubits; while these are mostly self--explanatory, the reader is referred to
standard texts \cite{gruska,nielsenchuang} for explanations of the notation.
For clarity, we describe the superdense coding in full below.

\begin{enumerate}
\item Two qubits, $P_{1}$ and $P_{2}$, are placed in an entangled state using
the Hadamard and $\mathsf{CNot}$ operations. Alice is given $P_{1}$, and Bob
is given $P_{2}$.

\item Alice selects a message, $(x,y)$, and applies the $i$th Pauli operator,
$\sigma_{i},$ to $P_{1}$, where $i=y+x(2+(-1)^{y})$. She transmits this
particle to Bob.

\item Bob applies the $\mathsf{CNot}$ gate from $P_{1}$ to $P_{2}$, and then
he applies the Hadamard gate to the former.

\item Bob measures the two particles, thus obtaining a pair of classical bits,
$(x^{\prime},y^{\prime})$. If no disturbance has occurred, this pair of bits
will match the original message, i.e. $(x^{\prime},y^{\prime})=(x,y)$.
\end{enumerate}

In what follows, we discuss the salient aspects of the \textsc{prism} model
for this protocol, which is listed in the appendix; this will serve as a
simple demonstration of the nature and structure of similar \textsc{prism}
models. The model of superdense coding consists of four \textsc{prism}
modules. Of these four, one module is generated automatically by
\textsc{prismgen} and describes the possible states of the two qubits. There
is a module specifying Alice's actions, and similarly one for Bob's. The
actions in these two modules correspond exactly to the successive application
of quantum gates in Fig. \ref{fig:dc-circ}.

Before we examine the workings of this model in detail, consider the following
observations, which highlight the capabilities of \textsc{prism}. In the
\textsc{prism} model, Alice's first action is to select one of the four
possible messages (represented by the integers 0, 1, 2, 3); each message has
an equal probability, $\frac{1}{4}$, of being chosen. This is an assumption we
made when constructing this model, but it is possible to specify different
respective probabilities for the four choices. Another point worth noting is
that, depending on which message is chosen, the protocol proceeds in one of
four distinct ways; \textsc{prism} actually considers \emph{all} these
possibilities when testing the validity of a property. This is precisely why
we advocate the use of model-checking for these analyses, as opposed to
simulation of quantum protocols, proposed elsewhere; simulators only treat one
of several possible executions at a time.

\begin{figure}
[ptb]
\begin{center}
\includegraphics[
height=1.4226in,
width=4.1303in
]%
{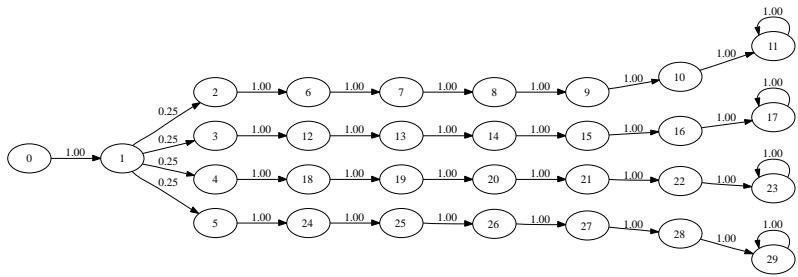}%
\caption[Internal probabilistic state transition system corresponding to the
\textsc{prism} model of superdense coding.]{Internal probabilistic state
transition system corresponding to the \textsc{prism} model of superdense
coding. }%
\label{fig-tree}%
\end{center}
\end{figure}

\textsc{prism} interprets the superdense coding model as the probabilistic
transition system\footnote{Note that the probabilities in this diagram arise
from Alice's choice of message, not from measurement outcomes. In general, it
is measurement that produces probabilistic transitions.} depicted in Fig.
\ref{fig-tree}. The nodes in the graph correspond to the internal state
numbers which \textsc{prism} assigns to each step in the protocol. Each
internal state number corresponds to a tuple with the states of all variables
in a particular model.%

%TCIMACRO{\FRAME{ftbpFU}{4.1303in}{1.4226in}{0pt}{\Qcb[Internal probabilistic
%state transition system corresponding to the \QTR{sc}{prism} model of
%superdense coding.]{Internal probabilistic state transition system
%corresponding to the \QTR{sc}{prism} model of superdense coding.
%}}{\Qlb{fig-tree}}{dc2-imp-pr-graph.ps}{\special{ language "Scientific Word";
%type "GRAPHIC";  maintain-aspect-ratio TRUE;  display "USEDEF";
%valid_file "F";  width 4.1303in;  height 1.4226in;  depth 0pt;
%original-width 13.6744in;  original-height 4.6492in;  cropleft "0";
%filename 'dc2-imp-pr-graph.ps';file-properties "XNPEU";}}}%
%BeginExpansion
%EndExpansion

Read from left to right, Fig. \ref{fig-tree} shows the succession of internal
state numbers for the four possible messages that Alice may send to Bob in
superdense coding, The initial state, labelled 0, is where all variables are
first set. In internal state 1, Alice makes her choice of message, setting the
\texttt{msg} variable accordingly and leading to one of the
internal\textsc{\ }states 2, 3, 4, or 5 with equal probability. The succession
of \textsc{prism}'s internal\textsc{\ }states in Table 1, which includes the
value of each variable in the model, corresponds to the case in which Alice
has chosen the message \textbf{(1, 0)} and, hence, applied the $\sigma_{1}$
operator to her qubit. The quantum state of the two-qubit system is
represented by the variable \texttt{state}, which corresponds to the actual
quantum state indicated in the final column of this table.%

%TCIMACRO{\TeXButton{Start Table}{\begin{table}[t]}}%
%BeginExpansion
\begin{table}[t]%
%EndExpansion

\begin{center}%
\begin{tabular}
[c]{|c|c|c|c|c|c|c|}\hline
{\small State number} & \texttt{alice\_step} & \texttt{msg} &
\texttt{bob\_step} & \texttt{result} & \texttt{state} & {\small Quantum
State}\\\hline
{\small 0} & {\small 0} & {\small 0} & {\small 0} & {\small 0} & {\small 0} &
$\left\vert 00\right\rangle $\\\hline
{\small 1} & {\small 1} & {\small 0} & {\small 0} & {\small 0} & {\small 8} &
$\frac{1}{\sqrt{2}}\left(  \left\vert 00\right\rangle +\left\vert
11\right\rangle \right)  $\\\hline
{\small 4} & {\small 2} & {\small 2} & {\small 0} & {\small 0} & {\small 8} &
{}$\frac{1}{\sqrt{2}}\left(  \left\vert 00\right\rangle +\left\vert
11\right\rangle \right)  $\\\hline
{\small 18} & {\small 3} & {\small 2} & {\small 0} & {\small 0} & {\small 9} &
{}$\frac{1}{\sqrt{2}}\left(  \left\vert 01\right\rangle +\left\vert
10\right\rangle \right)  $\\\hline
{\small 19} & {\small 3} & {\small 2} & {\small 1} & {\small 0} & {\small 9} &
{}$\frac{1}{\sqrt{2}}\left(  \left\vert 01\right\rangle +\left\vert
10\right\rangle \right)  $\\\hline
{\small 20} & {\small 3} & {\small 2} & {\small 2} & {\small 0} & {\small 7} &
$\frac{1}{\sqrt{2}}\left(  \left\vert 00\right\rangle -\left\vert
10\right\rangle \right)  ${}\\\hline
{\small 21} & {\small 3} & {\small 2} & {\small 3} & {\small 0} & {\small 2} &
{}$\left\vert 10\right\rangle $\\\hline
{\small 22} & {\small 3} & {\small 2} & {\small 4} & {\small 0} & {\small 2} &
{}$\left\vert 10\right\rangle $\\\hline
{\small 23} & {\small 3} & {\small 2} & {\small 5} & {\small 2} & {\small 2} &
{}$\left\vert 10\right\rangle $\\\hline
\end{tabular}

\end{center}%

%TCIMACRO{\TeXButton{End Table}{\caption{The transitions of the \textsc{prism}
%model for superdense coding for the case when the message chosen by Alice is \textbf
%{(1,0)}.}
%\label{prism-table}
%\end{table}}}%
%BeginExpansion
\caption{The transitions of the \textsc{prism}
model for superdense coding for the case when the message chosen by Alice is \textbf
{(1,0)}.}
\label{prism-table}
\end{table}%
%EndExpansion

When Bob has finished his measurement, and the dense coding protocol
terminates, one of the internal\textsc{\ }states 11, 17, 23, 29 is reached,
corresponding to the final nodes in the graph in Fig. \ref{fig-tree}. The
property required for verification must be expressed in terms of the final
state. When the dense coding protocol terminates, Bob's measurement result,
i.e. the pair of classical bits $(x^{\prime},y^{\prime})$, must match Alice's
original choice $(x,y)$. This requirement is expressed using \textsc{pctl,} as follows:%

\begin{equation}
\text{P}\geqslant\text{1 }[\text{ }\mathbf{true}\text{ }\mathcal{U}\text{
}((\text{\texttt{protocol\_finished}})\text{ }\wedge\text{ }%
(\text{\texttt{result}}=\text{\texttt{msg}}))\text{ }]\label{pctl-dense}%
\end{equation}

The \textsc{pctl }formula in (\ref{pctl-dense}) stipulates that the
probability of Bob's result matching Alice's choice is 1. Model--checking with
\textsc{prism} confirms that this property holds (i.e. this property is
\textbf{true} for all executions of the model). We have thus proven, using the
\textsc{prism} model--checker, that the dense coding protocol always succeeds
in transmitting two classical bits using a single qubit. Clearly, this is not
difficult to prove by hand; however, we have used dense coding as a simple
demonstration of our approach.

\subsection{Quantum Teleportation}

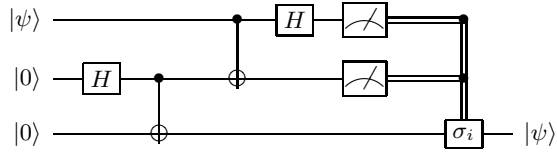
\begin{figure}[ht]
\centerline{\Qcircuit@C=1.2em @R=1em {
& \\
& \\
\lstick{\ket{\psi}} &\qw& \qw&\qw& \ctrl{1} & \gate{H} & \meter&\cw
& \control\cw& \\
\lstick{\ket{0}}     &\gate{H}  & \ctrl{1}&\qw&  \targ& \qw& \meter
&\cw& \control\cw\cwx& \\
\lstick{\ket{0}}     &\qw& \targ&\qw&  \qw& \qw&  \qw&\qw& \gate{\sigma_i}%
\cwx& \rstick{\ket{\psi}}\qw& \\
}}
\caption{Quantum circuit diagram for the teleportation protocol.}
\label{fig:tp-circ}
\end{figure}

Our next example is the quantum teleportation protocol \cite{Bennett93}, which
involves a computation on three qubits. Teleportation is a process that
exploits entanglement in order to transmit an arbitrary qubit state using only
a classical communications channel. The quantum circuit for teleportation is
shown in
%TCIMACRO{\TeXButton{TeX penalty}{\penalty10000} }%
%BeginExpansion
\penalty10000
%EndExpansion
Fig.\ref{fig:tp-circ}.%

%TCIMACRO{\TeXButton{Teleportation quantum circuit}{\begin{figure}[ht]
%\centerline{\Qcircuit@C=1.2em @R=1em {
%& \\
%& \\
%\lstick{\ket{\psi}} &\qw& \qw&\qw& \ctrl{1} & \gate{H} & \meter&\cw
%& \control\cw& \\
%\lstick{\ket{0}}     &\gate{H}  & \ctrl{1}&\qw&  \targ& \qw& \meter
%&\cw& \control\cw\cwx& \\
%\lstick{\ket{0}}     &\qw& \targ&\qw&  \qw& \qw&  \qw&\qw& \gate{\sigma_i}%
%\cwx& \rstick{\ket{\psi}}\qw& \\
%}}
%\caption{Quantum circuit diagram for the teleportation protocol.}
%\label{fig:tp-circ}
%\end{figure}}}%
%BeginExpansion
%EndExpansion

The \textsc{prism} model of teleportation is similar in appearance to that for
superdense coding, and it is not included here due to lack of space. It is a
transformation on a collection of three qubits, as opposed to the two for
superdense coding. This calls for the 3-qubit version of \textsc{prismgen}.
Other than this, the \textsc{prism} model itself is unremarkable, and matches
the structure of the quantum circuit for teleportation, given in the appendix.
Verifying the teleportation protocol with \textsc{prism} is more involved.
Short of manual calculation, it is not possible to predict what the quantum
state of the entire 3-qubit system will be at the end of the teleportation
protocol; indeed, there are several possible final states, depending on which
quantum state Alice chooses to transmit to start with. We are interested in
checking that the state of Bob's qubit matches Alice's original qubit state,
$\left\vert \psi\right\rangle $, which is assumed to be one of $\left\vert
0\right\rangle ,\left\vert 1\right\rangle ,\left\vert +\right\rangle
\left\vert -\right\rangle .$ To formulate a usable property for verification,
we need to express this requirement in terms of the overall state of the
3--qubit system.

Formally, the specification of the teleportation protocol is this: if the
initial state of the 3-qubit system is of the form $\left\vert \psi
\right\rangle \otimes\left\vert 00\right\rangle ,$ then the final state will
be of the form $\left\vert \phi\right\rangle \otimes\left\vert \psi
\right\rangle $, where $\left\vert \phi\right\rangle $ is a two--qubit state.
Let's consider this in more detail. If Alice chooses to teleport $\left\vert
\psi\right\rangle =\left\vert 0\right\rangle ,$ the final state of the
3--qubit system will be of the form $\left\vert \phi\right\rangle
\otimes\left\vert 0\right\rangle $. Similarly, if Alice chooses to teleport
$\left\vert \psi\right\rangle =\left\vert 1\right\rangle $, the final state of
all three qubits will be of the form $\left\vert \phi\right\rangle
\otimes\left\vert 1\right\rangle $. Finally, if Alice chooses to teleport the
superposition $\left\vert \psi\right\rangle =\frac{1}{\sqrt{2}}\left\vert
0\right\rangle +\frac{1}{\sqrt{2}}\left\vert 1\right\rangle $, the final state
of the three qubits will be of the form $\left\vert \phi\right\rangle
\otimes\left(  \frac{1}{\sqrt{2}}\left\vert 0\right\rangle +\frac{1}{\sqrt{2}%
}\left\vert 1\right\rangle \right)  .$

Clearly, the \textsc{pctl} property necessary for verification will depend on
the choice of $\left\vert \psi\right\rangle $; it will stipulate that, when
the teleportation protocol has completed, the final state of the 3-qubit
system will have one of the forms given above. In particular, if the input
state is $\left\vert 0\right\rangle ,$ the necessary property is
\begin{equation}
\text{P}\geqslant\text{1 }[\text{ }\mathbf{true}\text{ }\mathcal{U}\text{
}((\text{\texttt{telep\_end}})\text{ }\wedge\text{ }((\text{\texttt{st}}%
=s_{1})\text{ }\vee\text{ }\cdots\text{ }\vee\text{ }(\text{\texttt{st}}%
=s_{n})))\text{ }]\label{pctl-telep}%
\end{equation}

\noindent where \texttt{telep\_end} is a predicate which is \texttt{true} when
the protocol completes, and the values $s_{1},\ldots,s_{n}$ represent quantum
states of the form $\left\vert \phi\right\rangle \otimes\left\vert
0\right\rangle $. If the input state is $\left\vert 1\right\rangle ,$ the
necessary property has exactly the same form as (\ref{pctl-telep}), but the
values $s_{1},\ldots,s_{n}$ represent quantum states of the form $\left\vert
\phi\right\rangle \otimes\left\vert 1\right\rangle $; similarly for the case
when the input state is the superposition $\frac{1}{\sqrt{2}}\left\vert
0\right\rangle +\frac{1}{\sqrt{2}}\left\vert 1\right\rangle $.

In other words, in order to formulate the property needed to verify the
protocol, we need to choose the input states and determine the possible final
states of the three--qubit system in advance. This may be seen as begging the
question; there is little point in verifying a protocol whose final outcome
has already been calculated by hand. We have developed an auxiliary tool to
\textsc{prismgen}, which computes the internal state numbers $s_{1}%
,\ldots,s_{n}$ corresponding to the desired final states. When the
\textsc{pctl} property for a particular input is supplied to \textsc{prism},
the tool proves that the teleportation model works as expected. Since the
model--checker necessarily constructs a finite state space for the model, the
teleportation protocol can only be verified for a specific, known set of
inputs, rather than an arbitrary quantum state.

\subsection{Quantum Error Correction\label{sec:errorcorr}}

Our third and final example is the quantum bit--flip code for error correction
\cite{steane-err-corr}. In order to correct a single bit flip error, which may
occur during the transmission of a single qubit state, this code represents
the state by using a collection of three qubits. In particular, the qubit
state $\left\vert 0\right\rangle $ is encoded as $\left\vert 000\right\rangle
$ and the state $\left\vert 1\right\rangle $ is encoded as $\left\vert
111\right\rangle $. A bit flip error on the second qubit, for example,
transforms $\left\vert 000\right\rangle $ into $\left\vert 010\right\rangle $.

\begin{figure}[ht]
\centerline{\Qcircuit@C=1.3em @R=.6em {
\lstick{\ket{\psi}}               & \qw&\ctrl{1}         & \ctrl{2}
& \qw&\qw& \ctrl{3}           & \qw& \ctrl{4}          & \qw& \qw
& \multigate{2}{\sigma_{1, i}}  & \rstick{\ket{\psi}}\qw\\
&                            \lstick{\ket{0}}       & \targ& \qw& \qw&\qw
&\qw& \ctrl{2}         & \qw& \qw& \qw& \ghost{\sigma_{1, i}} \qw&   \qw\\
&                            \lstick{\ket{0}}       & \qw& \targ& \qw&\qw
& \qw& \qw& \qw& \ctrl{2}                & \qw& \ghost{\sigma_{1, i}}
\qw&   \qw\\
&                             &       &                         &                            & \lstick
{\ket{0}}                                                  & \targ& \targ
& \qw& \qw& \meter& \control\cw\cwx\\
&                              &      &                         &                            & \lstick
{\ket{0}}                                                  & \qw& \qw
& \targ&  \targ& \meter& \control\cw\cwx\\
}}
\caption{Quantum circuit diagram for the qubit bit--flip code.}
\label{fig:ec-circ}
\end{figure}
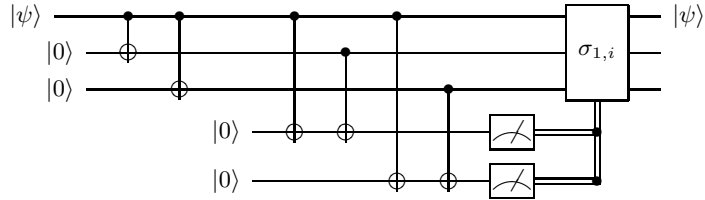

In order to detect such an error, two additional qubits are used; they are
known as \emph{ancillas}. By applying a sequence of operations and
measurements to the ancillas, the so--called \emph{error syndrome} is
obtained, which determines the location of the error. Then, the $\sigma_{1}$
operator is applied to the erroneous qubit, thus restoring the initial quantum
state of the 3--qubit system. The quantum circuit for the bit--flip code is
given in Fig. \ref{fig:ec-circ}.%

%TCIMACRO{\TeXButton{Quantum error corr circuit}{\begin{figure}[ht]
%\centerline{\Qcircuit@C=1.3em @R=.6em {
%\lstick{\ket{\psi}}               & \qw&\ctrl{1}         & \ctrl{2}
%& \qw&\qw& \ctrl{3}           & \qw& \ctrl{4}          & \qw& \qw
%& \multigate{2}{\sigma_{1, i}}  & \rstick{\ket{\psi}}\qw\\
%&                            \lstick{\ket{0}}       & \targ& \qw& \qw&\qw
%&\qw& \ctrl{2}         & \qw& \qw& \qw& \ghost{\sigma_{1, i}} \qw&   \qw\\
%&                            \lstick{\ket{0}}       & \qw& \targ& \qw&\qw
%& \qw& \qw& \qw& \ctrl{2}                & \qw& \ghost{\sigma_{1, i}}
%\qw&   \qw\\
%&                             &       &                         &                            & \lstick
%{\ket{0}}                                                  & \targ& \targ
%& \qw& \qw& \meter& \control\cw\cwx\\
%&                              &      &                         &                            & \lstick
%{\ket{0}}                                                  & \qw& \qw
%& \targ&  \targ& \meter& \control\cw\cwx\\
%}}
%\caption{Quantum circuit diagram for the qubit bit--flip code.}
%\label{fig:ec-circ}
%\end{figure}}}%
%BeginExpansion
%EndExpansion

\noindent For the diagram we have assumed that a bit--flip error does occur
prior to the computation of the syndrome.

Our \textsc{prism} model of the protocol for the quantum bit--flip code
includes a channel which perturbs the transmitted qubit with a chosen
probability; this probability is a parameter of the model, and can be varied
as required. The model uses the output from the 3--qubit version of the
\textsc{prismgen} tool. When the syndrome computation is taken into account,
there are in total five qubits whose states need to be modelled; since we have
not yet implemented a code generator for 5--qubit quantum systems, the state
transitions for the syndrome computation are calculated in advance and
manually coded into \textsc{prism}.

To verify the correctness of the quantum bit--flip code, we need to ensure
that: independently of which of the three qubits is perturbed and with what
probability this occurs, the protocol does succeed in correcting the
error.\ Thus, at the end of the protocol, the state of the 3--qubit system
should be in one of the following forms (where $\left\vert \phi\right\rangle $
is a two--qubit state):%

\begin{align}
\left\vert 0\right\rangle \otimes\left\vert \phi\right\rangle ,\text{ }  &
\text{if the input state was }\left\vert 0\right\rangle \label{form0**}\\
\left\vert 1\right\rangle \otimes\left\vert \phi\right\rangle ,\text{ }  &
\text{if the input state was }\left\vert 1\right\rangle \label{form1**}\\
\left(  \frac{1}{\sqrt{2}}\left\vert 0\right\rangle +\frac{1}{\sqrt{2}%
}\left\vert 1\right\rangle \right)  \otimes\left\vert \phi\right\rangle
,\text{ }  & \text{if the input state was }\frac{1}{\sqrt{2}}(\left\vert
0\right\rangle +\left\vert 1\right\rangle )\label{form0**+1**}%
\end{align}

The properties used in \textsc{prism} to verify the protocol are analogous to
those for teleportation, taking the form%

\begin{equation}
\text{P}\geqslant\text{1 }[\text{ }\mathbf{true}\text{ }\mathcal{U}\text{
}((\text{\texttt{qbf\_end}})\text{ }\wedge\text{ }((\text{\texttt{st}}%
=s_{1})\text{ }\vee\text{ }\cdots\text{ }\vee\text{ }(\text{\texttt{st}}%
=s_{n})))\text{ }]
\end{equation}

\noindent where \texttt{qbf\_end} is a predicate which holds when the protocol
completes, and the values $s_{1},\ldots,s_{n}$ represent quantum states of one
of the forms given in (\ref{form0**})--(\ref{form0**+1**}). \textsc{prism}
confirms that the protocol does indeed leave the 3--qubit system in one of
these forms, depending on the input, as expected.

\section{Challenges and Future Prospects}

We have demonstrated our approach to the analysis of quantum communication
protocols using three simple examples. There is significant scope for future
work, ranging from improvements to our current code--generation techniques, to
the automated verification of larger systems, such as quantum cryptographic devices.%

%TCIMACRO{\TeXButton{State--Space Construction Techniques}{\paragraph
%{State--Space Construction Techniques.}} }%
%BeginExpansion
\paragraph{State--Space Construction Techniques.}
%EndExpansion
At present we explicitly construct state spaces and transition tables for
systems involving up to three qubits and the $H$, $\mathsf{CNot}$ and
$\sigma_{i}$ operators. We have informally reached the conclusion that, for
any number of qubits, there is a finite set of states which is closed under
these operators. It is not directly obvious how many states these are, but
this could be established computationally. There is a mathematical framework
called the stabilizer formalism , which could be used to calculate these
states. Stabilizer circuits, which only include Clifford group gates
(Hadamard, $\mathsf{CNot}$ and the \emph{phase gate}
\cite{gruska,nielsenchuang}) and single--qubit measurements, are important in
quantum error correction and fault-tolerant computation. Furthermore, the
Gottesman--Knill theorem states that stabilizer circuits can be efficiently
simulated by a classical computer \cite{aaronson}. We conjecture that there is
a close correspondence between the Gottesman--Knill Theorem and the
calculation of the closed set of states outlined here (note that the Pauli
operators are derivable from the Clifford group gates). It is perhaps not
surprising that there should be a connection between a class of quantum
systems which can be efficiently simulated by a classical computer, and the
class of quantum systems which can be effectively model-checked.

We are also investigating another approach, based on a direct representation
of the coefficients in a quantum state as \textsc{prism} variables. By not
normalizing the states in the conventional way, the need for real numbers can
be avoided: for example, the state $\frac{\sqrt{3}}{2}|0\rangle+\frac{1}%
{2}|1\rangle$ would be represented by storing integer multiples of the
coefficients. The correct normalization factor would be applied when
calculating the probabilities for the measurement transitions. By storing
coefficients and calculating the probabilities explicitly within a
\textsc{prism} model, we would avoid the commitment in advance to a state
space of a particular structure. Initial attempts to do this have indicated
that it is not straightforward to implement with \textsc{prism}: to support
general coefficients with a reasonable precision, the range of values of the
\textsc{prism} variables must be large, and this leads to difficulties with
generating the state space.%

%TCIMACRO{\TeXButton{High--Level Modelling Languages}{\paragraph
%{High--Level Modelling Languages.}} }%
%BeginExpansion
\paragraph{High--Level Modelling Languages.}
%EndExpansion
The guarded transitions of \textsc{prism}'s modelling language make it awkward
to express some basic control structures such as sequencing. Each
\textsc{prism} module typically requires a variable which acts as a program
counter and must be explicitly incremented in each transition. We intend to
develop automatic translations from the high--level process calculus
\textsc{cqp} \cite{Nagarajan04} into \textsc{prism}'s native language.
Combining such a specification formalism for protocol models with a logic for
defining properties will allow us to verify quantum protocols at a higher
level. Ultimately we would like to make use of existing work on quantum logic,
for example \cite{MateusP:reaqs}.%

%TCIMACRO{\TeXButton{Modelling Larger Systems}{\paragraph
%{Modelling Larger Systems.}} }%
%BeginExpansion
\paragraph{Modelling Larger Systems.}
%EndExpansion
Our ultimate aim is to construct models of larger systems which combine
quantum and classical components, or which combine more than one quantum
protocol. For example, we are working on augmenting an existing model
\cite{nikosthesis} of the BB84 key--distribution protocol with descriptions of
authentication, secret--key reconciliation, and privacy amplification
protocols \cite{gruska}. As \textsc{prism} allows probabilities of particular
events to be calculated directly, we can obtain numerical values of
probability, such as those that arise in mathematical analyses of security; we
have taken advantage of this capability in our existing model of BB84. More
complex protocols generally involve larger numbers of qubits, leading to ever
greater state spaces for verification. However, it will often be the case that
the quantum state of a large system can be separated into independent parts.
By using a high--level modelling language such as \textsc{cqp}, we will be
able to identify non--interacting parts of the quantum state by static analysis.

\section{Conclusions}

We have established, for the first time, techniques for analyzing and
verifying quantum communication systems. Our key contributions are the
development of a general approach to modelling the state space of systems of
several qubits, and the introduction of techniques for defining properties of
quantum protocols in the logic \textsc{pctl}. We have illustrated our approach
by modelling and verifying three example protocols (superdense coding, quantum
teleportation, and quantum error correction) using \textsc{prism}. Although
these examples are simple, they are important building blocks of the theory of
quantum communication. Our approach is conceptually simpler than existing
mathematical proofs, and allows us to disambiguate protocol definitions and
assess their properties. What is more, it is easy to investigate the effect of
varying specific parameters and details once a model of a protocol has been
built. Having established fundamental and general techniques for formal
verification of quantum protocols, we are in a strong position to carry out
end--to--end verifications of larger systems, such as those used for quantum cryptography.

%\bibliographystyle{acm}
%\bibliography{../Bibliographies/concur05-biblio,../Bibliographies/fromMscThesis}

\section{Appendix}

\subsection{PRISM Model of a Single Qubit}
\begin{verbatim}
probabilistic
module Qubit
state : [0..3];  // 0 is |0>, 1 is |1>, 2 is |+>, 3 is |->
result : [0..1]; // Result of measurement in standard basis
[hadamard] (state=0) -> (state'=2);
[hadamard] (state=1) -> (state'=3);
[hadamard] (state=2) -> (state'=0);
[hadamard] (state=3) -> (state'=1);
[measure] (state=0) -> (state'=0)&(result'=0);
[measure] (state=1) -> (state'=1)&(result'=1);
[measure] (state=2) -> 0.5 : (state'=0)&(result'=0) 
                     + 0.5 : (state'=1)&(result'=1);
[measure] (state=3) -> 0.5 : (state'=0)&(result'=0) 
                     + 0.5 : (state'=1)&(result'=1);
endmodule
\end{verbatim}

\subsection{PRISM Model of Superdense Coding}
\begin{verbatim}
probabilistic 
//initialize to 1/sqrt2 (|00>+|11>), which is state=8
const int INITSTATE = 8; 
formula AL_FIN = (alice_step=3);
formula BOB_WAIT = (bob_step=0);
module Alice
alice_step : [0..3];
msg : [0..3];
// 1="00", 2="01", 3="10", 4="11"
[initialize]   BOB_WAIT &(alice_step=0) -> (alice_step'=1);
[]             BOB_WAIT &(alice_step=1) -> 
                          1/4 : (msg'=0) & (alice_step'=2) 
                        + 1/4 : (msg'=1) & (alice_step'=2) 
                        + 1/4 : (msg'=2) & (alice_step'=2) 
                        + 1/4 : (msg'=3) & (alice_step'=2); 
[sigma0] BOB_WAIT & (alice_step=2) & (msg=0) -> (alice_step'=3);
[sigma1] BOB_WAIT & (alice_step=2) & (msg=1) -> (alice_step'=3);
[sigma3] BOB_WAIT & (alice_step=2) & (msg=2) -> (alice_step'=3);
[sigma2] BOB_WAIT & (alice_step=2) & (msg=3) -> (alice_step'=3);
[sendtoBob] BOB_WAIT &(alice_step=3) -> (alice_step'=3);
endmodule
 
module Bob
bob_step : [0..5];
result : [0..3];
// 1="00", 2="01", 3="10", 4="11"
[sendtoBob]   AL_FIN & (bob_step=0) -> (bob_step'=1);
[cnot]        AL_FIN & (bob_step=1) -> (bob_step'=2);
[hadamard1]   AL_FIN & (bob_step=2) -> (bob_step'=3);
[measure]     AL_FIN & (bob_step=3) -> (bob_step'=4);
[]            AL_FIN & (bob_step=4) -> 
                   (result'=outcome) & (bob_step'=5);
[]            AL_FIN & (bob_step=5) -> (bob_step'=5); 
//End of protocol
endmodule
// ... 
// Automatically generated module with sigma0,sigma1,... 
// actions not included here due to its repetitivity 
// and the lack of space.
\end{verbatim}

\end{document}